\documentclass[prd,groupedaddress,nofootinbib,showpacs,12pt]{revtex4}

\usepackage{graphicx}
\usepackage{dcolumn}
\usepackage{bm}

\begin{document}

\title{Isometries of a D3-brane space}

\author{Henrique Boschi-Filho}
\email{boschi@if.ufrj.br}
\affiliation{Instituto de F\'{\i}sica, 
Universidade Federal do Rio de Janeiro, Caixa Postal 68528, RJ 21945-970 
-- Brazil}
\author{Nelson R. F. Braga}
\email{braga@if.ufrj.br}
\affiliation{Instituto de F\'{\i}sica,
Universidade Federal do Rio de Janeiro, Caixa Postal 68528, RJ 21945-970 
-- Brazil}

 
\begin{abstract} 
We obtain the Killing equations and the corresponding infinitesimal isometries for 
the ten dimensional space generated by a large number of coincident D3-branes. 
In a convenient limit this space becomes an
$AdS_5\times S^5$ which is relevant for the AdS/CFT correspondence. 
In this case, using Poincar\'e coordinates, we also write down the 
Killing equations and infinitesimal isometries. 
Then we obtain a simple realization of the isomorphism between $AdS$ isometries
and the boundary conformal group.

\end{abstract}


\maketitle

\section{Introduction}
 
In the supergravity approximation for low energy string theory one finds
non trivial background solutions known as Dp-branes\cite{HS}.
These objects were later identified as solitons in string theory\cite{Po}. 
The ten dimensional geometry generated by a large number of 
coincident D3-branes has a limit corresponding to an $AdS_5\times S^5$ space. 
This limit is an essential ingredient for the
duality found by Maldacena between string theory in $AdS$ spaces and
supersymmetric conformal field theories on their boundaries\cite{Malda1}
($AdS/CFT$ correspondence). 
Prescriptions detailing this correspondence
were presented in \cite{GKP,Wi} where it was shown how to calculate
boundary correlation functions from the $AdS$ bulk string theory 
(for a review and a wide list of references see\cite{Malda2}). 
Furthermore it was shortly pointed out \cite{Wi} that this duality can be understood 
as a realization of the holographic principle\cite{HOL1,HOL2,HOL3}. 
This principle states that in a quantum theory with gravity 
(as is the case of string theory) all the information contained in some spatial 
region can be mapped on a corresponding boundary. This principle was inspired by the 
study of quantum aspects of black hole entropy\cite{BHE1,BHE2}.

In this article we are going to study the isometries of the ten dimensional
space generated by D3-branes.
We will also consider the Killing equations and the corresponding isometries 
in the  $AdS_5\times S^5$ limit.
As it is well known $AdS_{n+1}$ space is maximally symmetric  
with isometry group $SO(2,n)$ which is isomorphic to the conformal group
defined in an $n$ dimensional flat space. 
Here we obtain a simple explicit realization of this isomorphism using Poincar\'e 
coordinates. These coordinates are important in the study of the $AdS/CFT$ 
correspondence, as for example in calculating boundary correlation functions.
Previous studies on the isometries of $AdS$ or asymptotically $AdS$ spaces
include \cite{IA1,GKP,Pe,IA2,IA3}. Also some aspects of the compactification
of  $AdS$ space and its relation to the D3-brane space were 
discussed in \cite{BB1}.

For a general space with coordinates $x^\mu $,  infinitesimal isometries 
are defined by the condition of invariance of the  metric $g_{\mu\nu}$
under the transformations\cite{CA}   
\begin{equation}
x^{\prime\mu} = x^\mu + \epsilon \xi^\mu\,,
\end{equation} 
                   
\noindent where $\epsilon\,$ is an arbitrary infinitesimal parameter. 
This implies the Killing equations
\begin{equation}
\nabla_\mu \xi_\nu +  \nabla_\nu \xi_\mu  \,=\,0
\end{equation}
\noindent or explicitly 
\begin{equation}
\label{KE}
\partial_\mu \xi_\nu +  \partial_\nu \xi_\mu - 2 \Gamma^\sigma_{\mu\nu} 
\xi_\sigma \,=\,0\,\,,
\end{equation}

\noindent where the Christoffel symbols are given by 

$$
\Gamma^\sigma_{\mu\nu} \,=\, { 1 \over 2} g^{\sigma\rho} \,\,\Big[ 
\,\,{\partial g_{\mu\rho}\over \partial x^\nu} + \,
{\partial g_{\nu\rho}\over \partial x^\mu} -\,
{\partial g_{\mu\nu}\over \partial x^\rho}\,\, \Big] \,\,.
$$

\section{D3-Brane space isometries}

The invariant measure $ds^2 = g_{\mu\nu} dx^\mu dx^\nu $ of the ten 
dimensional geometry generated by a large number $N$ of coincident D3-branes 
can be written as\cite{HS,GKP}  
\begin{equation}
\label{branemetric}
ds^2 \,=\, \Big( 1 + {R^4\over r^4} \Big)^{-1/2} ( -dt^2 + d{\vec x}^2 ) +  
\Big( 1 + {R^4\over r^4} \Big)^{1/2} (dr^2 + r^2 d\Omega^2_5 )
\end{equation}

\noindent where $R^4 \,=\, N/ 2\pi^2 T_3$ and $T_3$ is the tension of a single D3-brane. 
Changing the axial coordinate according to: $ z = R^2/r$, the metric takes the form:

\begin{equation}
\label{branemetricz}
ds^2 \,=\, {R^2\over z^2} \sqrt{ f(z)} \Big(\, -dt^2 + d{\vec x}^2 \,\Big) +  
 {R^2\over \sqrt{f(z)}} \Big(\, \,{dz^2\over z^2} + 
 d\Omega^2_5 \,\Big)\,\,,
\end{equation}

\noindent where 
$$f(z) =  {z^4\over z^4 + R^4} \,.$$

\noindent  From now on we take the Euclidean version of this metric so that
$(t, \vec x ) \equiv x^i\,\,,\, i = 1,..., 4$. The coordinates of $d\Omega_5$
are represented as $\theta^\alpha$ with $\alpha = 1,...,5$, assumed to be orthogonal:
$d\Omega_5^2 \,=\, {\tilde g}_{ \alpha \alpha } ( d \theta^\alpha )^2 $.
This implies the following non vanishing $z$ dependent Christoffel symbols
\begin{eqnarray} 
\label{G00}
\Gamma_{00}^0 &=& { f(z) - 2 \over z} \\
\Gamma_{0j}^i &=& -{ f(z) \over  z }\,\,\delta^i_j\\
\Gamma_{ij}^0 &=& { [f(z)]^2 \over z }\,\,{\bar \delta}_{ij}\\
\Gamma_{\alpha\beta}^0 &=& \, {\tilde g}_{[ \alpha \alpha ]} (\theta ) \,z \,
( 1 - f(z) )\,
\,{\bar \delta}_{\alpha\beta}\\
\label{GAA}
\Gamma_{0 \beta}^\alpha &=& {f(z) - 1  \over z }\,\,\delta_\beta^\alpha\,.
\end{eqnarray}

\noindent Our notation is:  $z \equiv x^0$ , Latin indices ($i,j,k,l ,...$) 
correspond to the variables $x^i $ and Greek indices ($\alpha, \beta, \gamma, ...$)
to the angular variables of  $d\Omega_5$.
Note that  $\Gamma_{\alpha\beta}^\gamma$ 
are in general non vanishing but are independent of $z$ (and $x^i$). 
The subscript $\,[\alpha]\,$ means that we are not summing over this index.
Note also that $\delta^i_j , \delta_\beta^\alpha $ are the usual Kronecker
tensors and we are defining the symbols ${\bar \delta}_{ij}$ and
$ {\bar \delta}_{\alpha\beta}$ to be one when their indices are equal
and zero otherwise, so that they are not tensors in this curved space-time.

For this D3-brane space the Killing equations (\ref{KE}) become
\begin{eqnarray}
\partial_i \,\xi_j &+& \partial_j \,\xi_i - { 2 [f(z)]^2 \over z }\,
{\bar \delta}_{ij}\, \xi_0 = 0\\
\partial_0\, \xi_i &+& \partial_i \,\xi_0 +  {2 f(z) \over  z }\, 
\xi_i = 0\\
\partial_0\, \xi_0 &+& { 2 - f(z) \over z}\, \xi_0 = 0\\
\partial_i \,\xi_\alpha &+& \partial_\alpha \,\xi_i = 0\\
\partial_0 \,\xi_\alpha &+& \partial_\alpha \,\xi_0 + \, 2 \,{ 1 - f(z) \over z} 
\,\xi_\alpha = 0\\
\partial_\alpha \,\xi_\beta &+& \partial_\beta \, \xi_\alpha +  
2\, z \,(\, f(z) - 1 \,) \,{\bar \delta}_{\alpha\beta} \, F_{\,[\alpha]} \,(\theta )\, 
 \xi_0 - \,2 \Gamma_{\alpha\beta}^\gamma \,\,\xi_\gamma = 0\,.
\end{eqnarray}

\noindent It is simpler to solve the corresponding equations for 
the contravariant Killing vectors that read
\begin{eqnarray}
{\bar \delta}_{jk} \partial_i \,\xi^k &+& {\bar \delta}_{ik} \partial_j \,\xi^k 
- { 2 f(z) \over z }\,
{\bar \delta}_{i\,j}\, \xi^0 = 0
\label{kx0}\\
{\bar \delta}_{ik} \partial_0\, \xi^k &+& {1 \over  f(z) } \partial_i \,\xi^0  = 0
\label{kxi}\\
\partial_0\, \xi^0 &+& { f(z) - 2  \over z}\, \xi^0 = 0\label{00}\\
{\bar \delta}_{ik} \,\partial_\alpha \,\xi^k 
 &+& {z^2\over f(z)} {\tilde g}_{\alpha\beta } \partial_i \,\xi^\beta
= 0\label{cr}
\\\partial_\alpha \,\xi^0 &+& 
z^2 {\tilde g}_{\alpha\beta } \partial_0 \,\xi^\beta  = 0 \label{kxa}\\
\partial_\sigma {\tilde g}_{\alpha\beta} \,\,\xi^\sigma &+& 
 {\tilde g}_{\beta \sigma} \partial _\alpha \xi^\sigma \,+\,
{\tilde g}_{\alpha \sigma} \partial _\beta \xi^\sigma \,
+\,2\,  \,{(\, f(z) - 1 \,)\over z} 
\,{\bar \delta}_{\alpha\beta} \, 
{\tilde g}_{[ \alpha\alpha ] } \,(\theta )\, 
 \xi^0 
= 0\,.
\end{eqnarray}

\noindent Equation (\ref{00}) constrains the dependence of $\xi^0$ on the coordinate 
$z$ to the form
\begin{equation}
\xi^0  \,=\, z\, [ f(z)]^{1/4} G ( x^i , \theta^\alpha )
\label{01}
\end{equation}

\noindent where $G ( x^i , \theta^\alpha )$ is some function to be determined. 
Now, equation (\ref{kx0}) takes the form
\begin{equation}
\label{gtil}
{\bar \delta}_{jk} \partial_i \,\xi^k \,+\, {\bar \delta}_{ik} \partial_j \,\xi^k 
\,=\, 2\, {\bar \delta}_{i\,j}  \,[f(z)]^{5/4} 
\, G  ( x^i, \theta^\alpha )    
\end{equation}
  
\noindent Taking $ i = j$ in the above equation we obtain 
$ [f(z)]^{5/4} G ( x^i, \theta^\alpha ) = \,(1/d)\, \partial_k \xi^k $
where $d = 4$ is the dimension of the space spanned by the coordinates $x^i$. 
Inserting this back into eq. (\ref{gtil}) we get

\begin{equation}
\label{eqgc}
{\bar \delta}_{jk} \partial_i \,\xi^k \,+\, {\bar \delta}_{ik} \partial_j \,\xi^k 
- {1\over 2 } {\bar \delta}_{i\,j}  \partial_k \xi^k   = 0
\end{equation}

\noindent which is the conformal group equation (for coordinates $x^i $).
From this equation we get: 
\begin{equation}
 \Big[ {\bar \delta}_{ij} \partial_k \partial_k  \,+\,2 \partial_i \partial_j \Big]
\partial_l \,\xi^l \,=\,0\,\,\,\,.
\end{equation}

\noindent This tells us that all the second derivatives of $\partial_l \,\xi^l $
with respect to the $x^i $ variables vanish. This determines the general 
quadratic form  of $\xi^i$ in the coordinates $x^j$, as usual in the conformal 
group transformations.
Then taking into account equation (\ref{gtil}) we find 
\begin{equation}
\label{xii}
\xi^i \,=\, x^k\omega^{ik} ( \theta^\alpha , z )  \,+\,
a^{i} ( \theta^\alpha , z ) \,+\,\Big[  x^i \lambda (\theta^\alpha ) + 
 x^j x^j d^i (\theta^\alpha )
- 2 x^i x^j d^j (\theta^\alpha ) \,\Big]\, { z^5 \over  ( R^4 + z^4 )^{5/4}}
\end{equation}

\noindent where $\omega^{ik} ( \theta^\alpha , z )\,=\,-\omega^{ki} 
( \theta^\alpha , z )\,$.  
Substituting this result in eq. (\ref{gtil}) we find 
$$ G ( x^i , \theta^\alpha ) = 
\lambda (\theta^\alpha ) - 2 x^j d^j (\theta^\alpha ) \,.$$

Now imposing equation (\ref{kxi})  we find 
\begin{eqnarray}
 x^k \partial_0\omega^{ik} ( \theta^\alpha , z )  &+&
\partial_0 a^{i} ( \theta^\alpha , z ) \,+\,\Big[  x^i \lambda (\theta^\alpha ) + 
 x^j x^j d^i (\theta^\alpha )
- 2 x^i x^j d^j (\theta^\alpha ) \,\Big]\,\partial_0 \Big(
 { z^5 \over  ( R^4 + z^4 )^{5/4}}\Big) \nonumber\\
&=&  2 z [ f( z) ]^{-3/4} d^i (\theta^\alpha ) \label{nee}
\end{eqnarray}

\noindent Comparing the terms with the same power of $z$ and $x^i$ we conclude that 
$$ \lambda = 0 $$
$$ \partial_0\omega^{ik} = 0$$
$$ d^i = 0 $$ 
$$ \partial_0 a^{i} = 0 \,\,,$$
 
\noindent so that $G ( x^i , \theta^\alpha ) = 0$. Then from
eqs. (\ref{01}) and (\ref{xii}) we find 
\begin{eqnarray}
\xi^0 &=& 0\\
\xi^i &=& x^k\omega^{ik} ( \theta^\alpha )  \,+\,
a^{i} ( \theta^\alpha )\,.
\end{eqnarray}

\noindent Using the result $\xi^0 = 0\,$ in eq. (\ref{kxa}) we get the condition:  
\begin{equation}
 \partial_0 \,\xi^\alpha\,=\,0
\label{v0}
\end{equation}

Now, since both  $\xi^\alpha $ and  $\,\xi^i $ are independent of the coordinate
$\,z\,$, equation (\ref{cr}) can only be satisfied if both terms vanish independently. 
This implies 
that: (i) $\omega^{ik} $ and  $ a^{i} $ are also independent of coordinates 
$\theta^\alpha$, that means they are {\it constants}; 
(ii) $\xi^\alpha $ are also independent of coordinates $x^i$.

So, the isometries of the ten dimensional D3-brane system can finally be written as
\begin{eqnarray}
\xi^0 &=& 0\\
\xi^i &=&  a^i + \omega^{ij} x^j \,\,\\
\xi^{\alpha} &=&  \xi^{\alpha} (\theta)\,.
\end{eqnarray}

 \noindent where $ \xi^{\alpha} (\theta) \equiv  \xi^{\alpha} (\theta^1,...,\theta^5)$
represent the usual isometries of $S^5$.

These solutions show that this space is not invariant for transformations 
in the coordinate $z\,$. This could be expected due to the presence of  
different factors of $ \,z\, $ in the D3-brane metric (\ref{branemetricz}). 
Note that the isometries in the $x^i$ coordinates correspond to  
ten independent parameters $a^i$ and $\omega^{ij}$.
These isometries are isomorphic to Poincar\'e transformations in 4 dimensional flat space.
Furthermore there is an $ S^5$ invariance in the $\theta^\alpha$ coordinates
with 15 independent parameters since $S^n$ spaces are maximally symmetric.

\section{AdS limit }

In the limit $ z >> R $  the D3-brane metric eq. (\ref{branemetricz})
takes the form of an $AdS_5\times S^5$ space  
\begin{equation}
\label{metric3}
ds^2=\frac {R^2 }{ z^2}\Big( dz^2 \,+(dx^i)^2\,
\Big)\,+\,R^2 d\Omega^2_5,
\end{equation}

\noindent  where the $AdS_5$ space with radius $R$ is represented 
by Poincar\'e coordinates ($ z, x^i$) with $i=1,...,4$. 
For this metric the Christoffel symbols are
\begin{eqnarray} 
\label{CF1}
\Gamma_{00}^0 &=& -{ 1\over z}\\
\Gamma_{0i}^j &=& -{ 1\over z}\,\,\delta^j_i \\
\Gamma_{ij}^0 &=& -{ 1\over z}\,\,{\bar \delta}_{ij}\\
\label{CF2}
\Gamma_{\alpha\beta}^0 &=& 0 \,=\, \Gamma_{0 \alpha}^\beta 
\end{eqnarray}

\noindent and $\Gamma_{\alpha\beta}^\gamma$  
are independent of $z $ and $x^i$ , 
as in the D3-brane space.
Note that the symbols (\ref{CF1})-(\ref{CF2}) can also be obtained from
the ones in the brane system eqs. (\ref{G00})-(\ref{GAA}) 
considering the limit $ z >> R $ such that $ f(z) \cong 1 - R^4/z^4 $. 

The Killing equations now take the form
\begin{eqnarray}
{\bar \delta}_{jk} \partial_i \,\xi^k &+& {\bar \delta}_{ik} \partial_j \,\xi^k 
- { 2  \over z }\,
{\bar \delta}_{i\,j}\, \xi^0 = 0
\label{ekx0}\\
{\bar \delta}_{ik} \partial_0\, \xi^k &+&  \partial_i \,\xi^0  = 0
\label{ekxi}\\
\partial_0\, \xi^0 &-&  { 1  \over z}\, \xi^0 = 0\label{e00}\\
{\bar \delta}_{ik} \,\partial_\alpha \,\xi^k 
 &+& z^2 {\tilde g}_{\alpha\beta } \partial_i \,\xi^\beta
= 0\label{e0a}
\\\partial_\alpha \,\xi^0 &+& 
z^2 {\tilde g}_{\alpha\beta } \partial_0 \,\xi^\beta  = 0 \label{e0i}\\
\partial_\sigma {\tilde g}_{\alpha\beta} \,\,\xi^\sigma &+& 
 {\tilde g}_{\beta \sigma} \partial _\alpha \xi^\sigma \,+\,
{\tilde g}_{\alpha \sigma} \partial _\beta \xi^\sigma \,= 0\,.
\end{eqnarray}

\noindent We follow the same procedure as in the previous section to find the isometries
in this case. From equation (\ref{e00}) we find 
\begin{equation}
\xi^0 \,=\,  z \,\, G ( x^i , \theta^\alpha )
\label{e01}
\end{equation}

\noindent In this case the solution for eq. (\ref{ekx0}) reads   
\begin{equation}
\label{exii}
\xi^i \,=\, x^k\omega^{ik} ( \theta^\alpha , z )  \,+\,
{\tilde a}^{i} ( \theta^\alpha , z ) \,+\,  x^i \lambda (\theta^\alpha ) + 
 x^j x^j d^i (\theta^\alpha )
- 2 x^i x^j d^j (\theta^\alpha )  
\end{equation}

\noindent where again $\omega^{ik} ( \theta^\alpha , z )\,=\,-\omega^{ki} 
( \theta^\alpha , z )\,$ and $ G ( x^i , \theta^\alpha ) = 
\lambda (\theta^\alpha ) - 2 x^j d^j (\theta^\alpha )\,$ as in the D3-brane case.

Now imposing equation (\ref{ekxi})  we find 

\begin{equation}
 x^k \partial_0\omega^{ik} ( \theta^\alpha , z )  \,+\,
\partial_0 {\tilde a}^{i} ( \theta^\alpha , z )  
=  2 z  d^i (\theta^\alpha ) \,.
\end{equation}

\noindent Note that this equation, in contrast to the corresponding
D3-brane equation (\ref{nee}), does not imply the vanishing of $\lambda$ and
$d^i$ (and consequently $\xi^0$) but only  
$$ \partial_0\omega^{ik} ( \theta^\alpha , z ) = 0$$
$$ \partial_0 {\tilde a}^{i}( \theta^\alpha , z ) = 2 z d^i ( \theta^\alpha )\,,$$
 
\noindent that means $ \omega^{ik} = \omega^{ik} ( \theta^\alpha) $ and
$  {\tilde a}^{i}( \theta^\alpha , z ) =  z^2 d^i ( \theta^\alpha )\, +\,
a^{i}( \theta^\alpha )$.

Then from eqs. (\ref{e0a}) and (\ref{e0i}) we see that $ \omega^{ik},\,d^i,\,
a^i $ and $\lambda $ do not depend on $\theta^\alpha$. Furthermore $\xi^\alpha$ can 
only depend on the $\theta$ coordinates. So the isometries in this case finally read
\begin{eqnarray}
\xi^0 &=& ( \lambda  - 2 d^i x^i )\, z \label{adsx0}\\
\xi^i &=&  a^i + \omega^{ij} x^j \,+\lambda x^i + d^i x^j x^j 
- 2 x^i x^j d^j \,+\,  z^2\,d^i  \,\,\label{adsxi}\\
\xi^{\alpha} &=&  \xi^{\alpha} (\theta)\label{xa}.
\end{eqnarray}

The 15 independent parameters $\,a^i,\, \omega^{ij}, \,d^i, \,\lambda\,$ 
of eqs. (\ref{adsx0}) and (\ref{adsxi}) represent the invariances of the $AdS_5$ space.  
This enlarges the isometries with respect to 
the D3-brane space. This happens due to the non vanishing $\xi^0$ in opposition to 
the previous case. The $S^5$ isometries represented by (\ref{xa}) are the same 
as in the D3-brane case. 
Note that the isometries of the $AdS_5$ and $S^5$ are independent as expected. 
Furthermore, the $AdS_5$ and $S^5$ spaces are both maximally symmetric 
so the isometry group of metric (\ref{metric3}) has 30 independent
parameters. 
 
Using these results, we now analyze the relation between the $AdS$ isometries 
and the induced isometries on surfaces of constant $z$.
First we consider a surface  $z = constant \ne 0 $.
The subgroup of isometries in this case corresponds to those with
$\xi^0 = 0$ which implies $\lambda = 0$ and $d^i = 0$.
Then in this case we find Poincar\'e invariance  

\begin{equation}
\xi^i =  a^i + \omega^{ij} x^j \,
\end{equation}

\noindent in such a surface. Additionally these surfaces 
also have $S^5$ isometries. 
So we see that the isometries of $AdS_5 \times S^5 $ at a fixed $z \ne 0 $ 
correspond exactly to the ones of the ten dimensional D3-brane system.

The very important particular case of the surface $ z = 0$ corresponds 
to the  $AdS$ boundary. 
On this surface the condition $\xi^0 = 0$ does not imply 
the vanishing of $\lambda$ and $d^i$, as can be seen from 
equation (\ref{adsx0}). These parameters remain arbitrary
so the isometries on the boundary are 

\begin{equation}
\label{conforme}
\xi^i =  a^i + \omega^{ij} x^j \,+\lambda x^i + d^i x^j x^j 
- 2 x^i x^j d^j\,\,.
\end{equation}

\noindent These isometries correspond to infinitesimal 
conformal transformations on the $x^i$ coordinates.

Let us now comment on the relations between the isometry subgroups in the bulk
and on the boundary. 
First note that dilatations on the boundary, defined by $\lambda \ne 0$ 
and the other parameters vanishing in eq. (\ref{conforme}) 
correspond to dilatations in the five dimensional $AdS$ space coordinates
given by eqs.(\ref{adsx0}),(\ref{adsxi}) with the same choice of parameters. 
Second, translations and rotations on the boundary given by non vanishing 
$a^i $ and $\omega^{ij}$ imply the same transformations in the $AdS$ bulk
with fixed value of the $z$ coordinate. Finally special conformal transformations
on the boundary given by non vanishing  $d^i$ correspond to the bulk isometries

\begin{eqnarray}
\xi^0 &=& - 2 d^i x^i \, z \\
\xi^i &=&   d^i\,( x^j x^j  + z^2 )\,
-\, 2 x^i x^j d^j \,\,.
\end{eqnarray}

\noindent This represents a special conformal transformation in the 
coordinates $x^\mu = ( z , x^i) $ with parameter $ d^\mu = (0 , d^i) $.

In summary, we found the isometries of the D3-brane space-time.
Then we considered the $AdS$ limit where we have shown explicitly 
that the isometry group acts on 
the boundary as the conformal group. This result is a simple 
realization in Poincar\'e coordinates of the isomorphism between 
these two groups. This is an essential ingredient for the $AdS/CFT$ correspondence.

\section*{Acknowledgments} 
We would like to thank Mauricio Calv\~ao for reading the manuscript
and for interesting discussions. 
The authors are partially supported by CNPq, FINEP , CAPES and FAPERJ 
- Brazilian research agencies.



\end{document}